\documentclass[twocolumn,preprintnumbers,aps,showpacs]{revtex4}
\def\bea{\begin{eqnarray}}
\def\eea{\end{eqnarray}}
\def\ben{\begin{equation}}
\def\een{\end{equation}}
\def\benu{\begin{enumerate}}
\def\enu{\end{enumerate}}

\def\sss{\scriptscriptstyle\rm}

\def\1var{(\bx_1...\bx\N)}

\def\bp{{\bf p}}
\def\br{{\bf r}}

\def\bx{{x}}
\def\by{{y}}

\def\bj{{\bf j}}

\def\c{_{\sss C}}
\def\s{_{\sss S}}
\def\xc{_{\sss XC}}

\def\N{_{\sss N}}
\def\H{_{\sss H}}

\def\ext{_{\rm ext}}

\def\sph_int{ {\int d^3 r}}

\input psfig.sty
\input rotate
\usepackage{graphicx}
\usepackage{psfrag}
\usepackage{amsmath}
\usepackage{bm}

\usepackage{epsfig}

\input epsf

\begin{document}
\title{Phase-Space Explorations in Time-Dependent Density Functional Theory
}
\date{\today}
\author{A. K. Rajam}
\affiliation{Department of Physics, The Graduate Center of the City University of New York, 365 Fifth Ave, New York, NY 10016, USA}
\author{Paul Hessler}
\affiliation{}
\author{Christian Gaun}
\affiliation{Department of Physics and Astronomy, Hunter College and City University of New York, 695 Park Avenue, New York, NY 10065, USA}
\author{Neepa T. Maitra}
\affiliation{Department of Physics, The Graduate Center of the City University of New York, 365 Fifth Ave, New York, NY 10016, USA}\affiliation{Department of Physics and Astronomy, Hunter College and City University of New York, 695 Park Avenue, New York, NY 10065, USA}

\begin{abstract}
We discuss two problems which are particularly challenging for
approximations in time-dependent density functional theory (TDDFT) to
capture: momentum-distributions in ionization processes, and
memory-dependence in real-time dynamics.  We propose an extension of
TDDFT to phase-space densities, discuss some formal aspects of such a
``phase-space density functional theory'' and explain why it could
ameliorate the problems in both cases. For each problem, a
two-electron model system is exactly numerically solved and analysed
in phase-space via the Wigner function distribution. 
\end{abstract}
\maketitle
\section{Introduction}
Time-dependent density functional theory (TDDFT) is a remarkably
successful theory of many-body systems in time-dependent external
potentials~\cite{RG84,GDP96,TDDFTbook}. The analog of static density functional theory~\cite{HK64}, TDDFT
is based on the Runge-Gross proof of a one-to-one mapping between the external potential and the time-dependent density of electrons
evolving under it, for a
specified initial wavefunction~\cite{RG84}. 
%The Runge-Gross theorem then states that all properties
Because knowledge of the external potential specifies the many-body
Hamiltonian, all properties of the interacting electronic system can
be extracted from just its one-body density and the initial-state.  In practise,
TDDFT utilizes the Kohn-Sham (KS) scheme, where the time-dependent
density of an interacting system evolving in an external potential $v\ext(\br,t)$ is calculated from a fictitious noninteracting system of
fermions moving in an effective potential, the KS potential
$v\s({\bf{r}},t)$, that is defined to reproduce the density of the interacting system. 
The KS potential is written as the sum:
\ben
\nonumber
v\s[n;\Phi_0](\br,t)= v\ext(\br,t) + v\H[n](\br,t)
+v\xc[n;\Psi_0,\Phi_0](\br,t)
\label{eq:vs}
\een
where $n(\br,t)$ denotes the time-dependent density, and
$\Psi_0$($\Phi_0$) is the initial interacting(noninteracting)
wavefunction.  The second term, $v\H[n](\br,t) = \int
\frac{n(\br',t)}{\vert\br - \br'\vert}d^3r'$, is the classical Hartree
potential, while the third is the exchange-correlation (xc) potential
$v\xc$. This is unknown as a functional of the time-dependent density
and initial-states, and must be approximated in practise.
Armed with an approximation for $v\xc[n;\Psi_0,\Phi_0](\br,t)$, one then propagates the time-dependent KS equation:
\ben
i\partial_t \phi_i(\br,t) = (-\nabla^2/2 + v\s(\br,t))\phi_i(\br,t)
\een
finding the $N$ single-particle orbitals $\phi_i(\br,t)$, that
initially made up the initial KS determinant $\Phi_0$. 
(Atomic units,  $e^2 =\hbar = m_e = 1$ are used throughout this paper).
The density of
the interacting system is obtained from $n(\br,t) =
\sum_i^{N}\vert\phi_i(\br,t)\vert^2$, and the Runge-Gross theorem then
assures us that all properties of the true system can be extracted
from the orbitals. In the linear response regime, in which lie the
majority of TDDFT applications so far, excitation energies and
oscillator strengths are obtained from a perturbative formulation of
these equations~\cite{PGG96,C96}, leading to the matrix formulation that operates in most of the quantum chemistry codes today. 

Clearly the accuracy of TDDFT then depends on the approximation used for the xc kernel. Despite being known generally to have ``memory-dependence'' -- 
that is, $v\xc(\br,t)$ depends on the history $n(\br,t'),t'\le t$ and
on the initial-states, $\Psi_0$ and $\Phi_0$~\cite{MBW02,MB01,M05b} --  almost all
approximations used today neglect this, and bootstrap a ground-state
functional: that is, the instantaneous density is input into an
approximate ground-state functional, $v\xc^{\rm
  A}[n;\Psi_0,\Phi_0](\br,t) = v\xc^{\rm gs}[n(\br t)](\br t)$. The
superscript A denotes this ``adiabatic approximation''.
In the linear response regime~\cite{PGG96,C96}, adiabatic TDDFT has achieved an
unprecedented balance between accuracy and efficiency for the
calculations of excitations and response properties. For the first
time, one can compute the quantum mechanical spectrum of systems as
large as biomolecules (see eg. Refs.~\cite{FA02,SRVB09,MLVC03}), run coupled electron-ion
dynamics on interesting chemical reactions~\cite{TTRF08}, and study
molecular transport through nanostructures~\cite{KCBC08}.  
Excitations for which the adiabatic approximation fails have been recognized, 
and, in various stages of being corrected~\cite{TH00,M05c,MZCB04,MW08}.  
Sometimes the spatial long-ranged-ness of the true functional is important, and approximations more sophisticated than the usual semi-local ones are required~\cite{GB04b,KKP04,FBLB02,VK96}.

But TDDFT applies also for strong field dynamics, and this is the
regime it is particularly promising for, given that correlated
wavefunction calculations for more than two or three electrons in
strong-fields become prohibitively expensive~\cite{PMDT00}.  Compared
to calculations of spectra, in addition to an approximation for the xc
potential, now a new ingredient needs to be considered in the
calculation: {\it observable functionals}, i.e. the observables of
interest must be expressed as functionals of the KS wavefunction. The
RG theorem guarantees that all properties of the true system can be
obtained from the KS orbitals, as they themselves are implicit
density-initial-state-functionals, but how?  If the property is
related directly to the density, then no additional observable
functional is needed.  For example, in high-harmonic-generation it is
the dipole moment, ${\bf d} = \int n(\br,t)\br d^3r$ that is of
interest, and indeed TDDFT calculations have been successful for
high-harmonic generation spectra of a range of interesting
systems~\cite{BNZM03,CC01, NBU04}.
But if we are interested in, for example, measuring double-ionization
probabilities, we require the interacting pair-density as a functional
of the density (or KS orbitals), and this is highly
non-trivial~\cite{WB06}. Non-sequential double-ionization is a
fascinating problem that dogged TDDFT for several years~\cite{LL98,LK05,WB06}: TDDFT's
promise in capturing electron correlation made it attractive for this
problem, but it was soon realized that not only does one need to go beyond the usual semilocal GGA's for the xc functional in order to obtain the knee structure~\cite{LK05}, but using an uncorrelated KS expression for the pair-density yielded a knee that was too high.  (In
Ref.~\cite{WB06}, an adiabatic correlated expression, based on the
ground-state Perdew-Wang pair density~\cite{PW92} was shown to lower
the knee appropriately). 
Approximating a general observable simply by
the appropriate operator evaluated on the KS wavefunction can lead to
gross inaccuracies, due to the lack of correlation in the KS
wavefunction itself. It is important to remember that the KS wavefunction is designed to reproduce the exact interacting one-body density, but is {\it not}
supposed to be an approximation to the true wavefunction. 
Indeed another aspect of the same double-ionization problem 
 recently demonstrated this: Wilken and Bauer's
calculations of ion-momentum-recoil distributions in
double-ionization~\cite{WB07} showed that the KS momentum-distributions were drastically
wrong, displaying a single maximum instead of the characteristic two-hump
structure, and with a significantly  overestimated magnitude. 
Typically KS momentum distributions are not those of the true system, even 
if the exact KS orbitals were used, that is if the exact $v\xc$ was known for the problem, and used to generate the KS orbitals. 
Another example of this will be given in Sec.~\ref{sec:momentum}.

Returning to the question of the approximate
$v\xc[n;\Psi_0,\Phi_0](\br,t)$: the accuracy of the adiabatic
approximation for real-time dynamics gets mixed reviews. In
strong-field double-ionization, the xc potential appears not to be
significantly non-local in time in a wide range of cases, although it
depends on how the field is ramped on~\cite{TRK08}. Dynamics of two
electrons in parabolic wells, on the other hand, yield an exact
correlation potential that appear to depend strongly on the
history~\cite{HMB02,U06, UT06}. We shall revisit Ref.~\cite{HMB02} in
Sec.~\ref{sec:Hooke}. Although significant advances have been made in understanding~\cite{MB01,MBW02} and modelling~\cite{DBG97,T05,T05b,UT06,VUC97,KB05}
memory-dependence, it remains today a difficult problem. 

%A third challenge for TDDFT lies in certain electronic quantum control
%problems. With the advent of attosecond lasers comes the possibility
%of controlling dynamics on the electronic time-scale, and TDDFT would
%seem to be an especially useful theoretical tool for
%this~\cite{Bandrauk}.  However, in certain cases, we encounter an
%interesting and difficult challenge whose origin lies in the the very
%nature of Kohn-Sham: dealing with a non-interacting system means that
%the time-evolving KS system remains in a single-Slater determinant
%state throughout, even when describing an interacting system whose
%natural occupation numbers change significantly. For example, the KS
%system describes a singlet singly-excited state by a single-Slater
%determinant; which is a poor description of the true state, which needs 
%at minium, two Slater determinants to describe.*** We shall return to this
%problem in Sec.~\ref{sec:Qmcontrol}.

%Usual approximations for the xc potential are adiabatic and spatially
%semi-local, i.e. they depend on the density instantaneously in time
%and semilocally in space. Usual observable functionals for strong-field dynamics simply apply th
 One situation in which the usual approximations (spatially semi-local
 and adiabatic density dependence) do poorly, is when the true
 wavefunction fundamentally cannot be described by a single
 Slater-determinant. This is also true for the ground state where
 molecular dissociation curves are a notorious
 problem~\cite{TMM09,CGGG00,GLB96}: when an electron-pair bond
 dissociates, the true wavefunction develops a Heitler-London
 character, which in a minimal description requires two
 Slater-determinants. The KS description however operates via a single
 Slater-determinant. The exact xc potential allows for the fundamental
 difference in the KS and true wavefunctions by developing rather
 stark and unusual peak and step features, that are difficult to
 capture in approximations. In time-dependent problems, coefficients
 of the interacting wavefunction expanded in a basis of
 single-Slater-determinants may change in time dramatically: whereas
 at one time a single-Slater determinant may dominate, at a later
 time, two single-Slater determinants are essential. Such situations
 may happen in electronic quantum control problems (now becoming
 experimentally accessible with the advent of attosecond
 lasers)~\cite{CYB06}. Consider for example evolving the ground-state of the He
 atom (1s$^2$) to its first accessible excited state (1s2p). The KS
 ground state is a single-slater determinant composed of a
 doubly-occupied spatial orbital. This evolves under the KS
 Hamiltonian, which is a one-body evolution operator. The target
 single excitation is however a {\it double} Slater determinant, and
 therefore cannot be reached by any one-body operator. The exact KS
 system attains the density of the true target state using a single
 orbital, not two. This yields a very unnatural description of the
 true system; consequently, the KS potential develops rather unnatural
 features, difficult to model~\cite{MBW02,M05b}. The inability of KS
 methods to change occupation numbers leads to challenges in developing approximations. 

The two ingredients needed in a TDDFT calculation - approximations for 
the xc potential and the relevant observable functional - determine the
accuracy of the results. As discussed above, sometimes the simplest
approximations work well, but in other cases, it is more challenging to derive
suitable approximations.
In this paper, we explore a generalization of TDDFT based on the
phase-space density. That is, instead of taking the coordinate-space
density, $n(\br,t)$ as our basic variable, we consider taking the
density in phase-space, $w(\br,\bp,t)$: a quasi-probability
distribution which is somewhat related to  the
probability of finding an electron at position $\br$ with momentum
$\bp$ (see Sec.~\ref{sec:Wigner}).  The idea here is two-fold: First, as the basic variable has
more information, functionals of them may be simpler to approximate,
or equally, simple functionals of them may be more accurate than
simple functionals of the coordinate-density alone, while retaining the favorable system-size scaling property of density functional theories. 
Second, more observables are directly
obtained without needing additional observable-functionals, in
particular, those pertaining to one-body operators, such as 
momentum, kinetic energy.  As the
coordinate and momentum operators do not commute, there is no unique
phase-space density function, and we begin in Sec.~\ref{sec:Wigner} by
discussing the Wigner function, which turns out to be a particularly
useful choice for our purposes. We discuss the 1-1 mapping between
Wigner functions and potentials, for a given initial-state.  We
discuss also the formal equivalence of the Wigner function with
time-dependent density-matrix functional theory, which has seen a
recent resurgence of interest~\cite{PGB07,GBG08,PGGB07}.  Next, we return to TDDFT and explore the true and KS phase-spaces, and the implications of a phase-space based theory, for two examples pertaining to the challenges in TDDFT described above:
momentum distributions (Sec.~\ref{sec:momentum}) and 
history-dependence in the xc functional (Sec~\ref{sec:Hooke}).

\section{The 1-body Wigner Phase-space Density}
\label{sec:Wigner}
Quantum phase-space distributions have played an important role in the
development of quantum mechanics. Although a precise joint probability
distribution in position and momentum is impossible due to the
Heisenberg uncertainty principle, various quasi-probability
distributions have been defined, which contain position and momentum
distribution information in a way consistent with the uncertainty
principle. Such distributions are particularly useful in a
semiclassical context, relating the quantum states to the underlying
classical trajectories, and have been exploited extensively, for
example in quantum chaos and quantum optics.  Phase-space approaches
have however been largely, although not
entirely~\cite{GP86,GBP84,PRG86,GNCB06,CDG07,CG07} neglected in density
functional theories, where position plays a preferred role over
momentum.  Most recently~\cite{GNCB06,CDG07,CG07}, Gill and co-workers,
have developed and tested models for the ground-state correlation energy based on
various ``phase-space intracules'' which are essentially  different
contractions of the two-body reduced Wigner function. For example, there is the 
Omega intracule, which is a function of three variables; the distance between two electrons, the magnitude of the difference in their momentum vectors, and the angle between their position and momentum~\cite{CG07}. 
This is motivated by the
very physical idea that the correlation between two electrons depends
on their relative momentum as well as on their relation
distance~\cite{R99}. 

Wigner functions were
first introduced in 1932~\cite{W32,CZ83,HCSW84}, with the intent
of application to many-body systems.  The Wigner function
quasi-probability distributions are real but not positive
semi-definite; indeed the negative areas have been interpreted as an
indication of non-classical behaviour, from interference to tunneling
processes, and research in how to classify these areas continues
today~(see for example Refs.~\cite{Spekkens}).
 Although initially introduced to treat many-particle dynamics, the
 majority of applications of the Wigner function actually involve one-particle
 systems, especially for making quantum-classical correspondence. 
 It has been argued that the Wigner function is a
 particularly suitable approach to studying
 transport~\cite{J04,CBBF04,CBJ07} due to its setting within a semiclassical picture while yet being rigorously quantum mechanical. Ref.~\cite{CBJ07} develops a formalism
 based on reduced-Wigner-functions for indistinguishable fermionic
 systems accounting for Pauli-exchange (but not correlation).

For a system of $N$ identical particles, the reduced $k$-body Wigner phase
space density is defined as the Fourier transform of the $k$th-order
density matrix $\rho_k$:
\begin{widetext}
\ben
w_k(\br_1..\br_k,\bp_1..\bp_k,t)=\left(\frac{1}{2\pi}\right)^{3k} \int d^3y_1..d^3y_k e^{i{\sum_k{\bf p_k}\cdot {\bf y_k}}} \rho_k({\bf r_1}-{\bf y_1}/2..{\bf r_k}-{\bf y_k}/2, {\bf r_1}+{\bf y_1}/2..{\bf r_k}+{\bf y_k}/2;t)
\een
\end{widetext}
where the $k$-th order density matrix is $\rho_k({\bf r_1'}..{\bf r_k'}, {\bf r_1}..{\bf r_2};t) = \left(\begin{array}{c}N\\k\end{array}\right)\int \Psi^*(\br_1'..\br_N',t)\Psi(\br_1..\br_N,t)dr_{k+1}..dr_N$.
We consider here a density-functional theory based on the one-body
Wigner function $w(\br,\bp,t) \equiv w_1(\br,\bp,t)$ (henceforth
referred to as simply Wigner function):
\ben
w(\br,\bp,t) = \left(\frac{1}{2\pi}\right)^3\int d^3y\rho_1(\br-{\bf y}/2,\br + {\bf y}/2,t)e^{i\bp\cdot\by}
\label{eq:wigner}
\een
The one-body densities in coordinate-space and momentum-space are
directly obtained by integration:
\bea
n(\br,t) = \int w(\br,\bp,t) d^3p\\
\tilde n(\bp,t) = \int w(\br,\bp,t) d^3r
\label{eq:nrnp}
\eea
Several observations can immediately be made about such a phase-space 
density-functional theory (PSDFT) whose basic variable is the Wigner 
phase-space density (taking the role of  the coordinate-space density in usual density functional theories). 
First, the
theory is formally
equivalent to the recently developed time-dependent one-body density-matrix functional theory (1DMFT)~\cite{PGB07,GBG08,PGGB07}: $w$ and
$\rho_1$ contain exactly the same information, in different forms
related via a Fourier transform. 
%For some purposes, for example proofs (see Sec.~\ref{sec:1-1}),
%$\rho_1$ is easier to work with, but for other purposes, for example,
%visualization, semiclassical analyses, and possibly the development of
%approximations $w$ may prove to be the more natural variable. 
Second, as in the 1DMFT, there is no Kohn-Sham counterpart: the
Wigner-function of an interacting system cannot be reproduced by any
non-interacting system in a local potential. In a non-interacting
system, the one-body reduced density matrix is idempotent, while in an interacting system, it
cannot be. It is interesting that generally while a non-interacting
system may be found that reproduces the coordinate-space density of an
interacting system, and a different non-interacting system may be
found reproducing its momentum-space density, one cannot find a non-interacting system that reproduces its phase-space density. 
Third, the Wigner function directly gives the expectation value of any one-body
operator: no additional observable functional would be needed, for
example, for kinetic energy or momentum distributions.
Fourth, the one-to-one mapping between the potential and the phase-space density, for a given initial-state, follows directly from the Runge-Gross theorem:
We have
\bea w(r,p,t) \to n(r,t) \\ 
\left(n(\br,t),\Psi_0\right) \to \Psi(\br_1..\br_N,t) 
\eea
 where both arrows indicate a unique mapping, the first following from $p$-integration (Eq.~\ref{eq:nrnp}), 
and the second from RG, while also
\ben
\Psi(\br_1..\br_N,t) \to w(r,p,t)
\een 
uniquely for a specified initial-state. Hence, for local external
potentials, there is a 1-1 mapping between the one-body Wigner
function and the many-body wavefunction for a given initial state,
i.e. that all observables may be extracted from the one-body Wigner
function and initial-state alone.

We note that this 1-1 mapping holds for local external potentials
(i.e. purely multiplicative in coordinate-space. Gilbert~\cite{G75}
has shown that for the {\it ground state}, the extension of the
Hohenberg-Kohn therem to one-body reduced density matrices applies also to the wider class of {\it
  non-local} potentials, $v(\br,\br')$. Formally, one may be led to
search for a one-to-one $\rho_1 - v(\br,\br')$ mapping in the
time-dependent case also, given that these are conjugate variables
that couple together in an energy functional, (c.f. the $n-v$
mapping of Runge-Gross or Hohenberg-Kohn~\cite{HK64}, and the $A-j$ mapping of Ghosh and Dhara in TD-current-density
functional theory~\cite{GD88,VK96}.
Although a proof for time-dependent non-local potentials is yet to be found, 
the mapping may be generalized to vector potentials ${\bf A}(\br,t)$, extending the realm of application to include magnetic fields:
\bea
w \to \bj(\br,t)\\
\left(\bj(\br,t),\Psi_0\right) \to \Psi(\br_1..\br_N,t) 
\eea
where the first arrow follows from 
\ben
{\bf j}(\br,t) = \int d^3 {\bf p} w(\br,{\bf p},t)
\een
and the second from the Ghosh-Dhara
proof~\cite{GD88} of the 1-1 mapping between currents and
vector-potentials. That is, any observable of any time-dependent
interacting electronic system evolving under an external vector
potential, is a functional of its initial wavefunction and time-dependent
one-body phase-space density.

The Runge-Gross proof proceeded in two steps, where the first proved a
one-to-one mapping between currents and (scalar) potentials for a
given initial state.The second step, to go from currents to densities,
holds under the condition that a boundary term, that involves the density and the gradient of the potential, vanishes.
The implications of this boundary condition have been discussed in Ref.~\cite{GK90}; it means that TDDFT cannot be applied to periodic systems in uniform electric fields~\cite{MSB03}.   We point out that here there is no requirement 
on the boundary
conditions for the $w-v$ (or $w-{\bf A}$ mappings to hold.

%Our proof is simply a re-hash of the RG proof but applied instead to
%the one-body Wigner function. That is, we consider a fixed initial
%state $\Psi_0$ and ask, can we find two potentials $v$ and $\tilde v$
%that yield the same phase-space density evolution for all time,
%$w(\br,\bp,t) = \tilde w(\br,\bp,t)$? In the following we shall work
%with the density-matrix; by Eq.~\ref{eq:wigner}, the 1-1 mapping for
%one implies the 1-1 mapping for the other.
%**fill in from other notes made in Benasque**

%Notice, there are no boundary conditions required for the proof,
%unlike in TDDFT. This can be understood from TD-current-density-FT,
%where also there are no b.cs...**explain**

%T physical observables are, in principle, functionals of the
%time-dependent one-body phase-space distribution. 

%Generally, t-dep 1-1 holds for any order of DM ...

Time-dependent phase-space functional theory involves 
solving the equation of motion for the Wigner function:
\begin{widetext}
\bea
\nonumber
\dot {w}({\bf{r,p}},t)&=&\left[-{\bf{p}}\cdot\nabla  -i\int d^3p'\int d^3y e^{-i({\bf{p-p'}})\cdot{\bf{y}}}\left[v_{ext}({\bf{r+\frac{y}{2}}})-v_{ext}({\bf{r-\frac{y}{2}}})\right]\right]w({\bf{r,p}},t) \\
&-&i\int d^3yd^3p_1d^3p_2d^3r_2 e^{i(\bp-\bp_1)\cdot\by}\left(\frac{1}{\vert \br-\br_2+\frac{\by}{2}\vert} - \frac{1}{\vert \br-\br_2-\frac{\by}{2}\vert}\right)w_2(\br,\br_2,\bp_1,\bp_2,t)
\eea
\end{widetext}
where the last term must be approximated as a functional of the one-body $w$ and
the initial-state $\Psi_0$.
This equation follows directly from that for the density-matrix, which may be more familiar:
\begin{widetext}
\ben
i\dot{\rho_1}({\bf{r'}},{\bf{r}},t)=\left(-\nabla^2/2 + v\ext({\bf{r}},t)+ \nabla'^2/2 - v\ext({\bf{r'}},t)\right)\rho_1({\bf{r'}},{\bf{r}},t)+
\int d^3r_2\left(\frac{1}{|{\bf{r-r_2}}|}-\frac{1}{|{\bf{r'-r_2}}|}\right)\rho_2(\br',\br_2,\br,\br_2,t)
\een
\label{eq:rho1dot}
\end{widetext}
where $\rho_2$ is the second-order density-matrix, to be approximated
as a functional of $\rho_1$.  This equation is first in the BBGKY (Bogoliubov-Born-Green-Kirkwood-Yvon) heirarchy of
reduced density matrix evolutions, which gives the equation of motion
for the $k$th reduced DM in terms of the $(k+1)$st and lower DM's.
Eq.~\ref{eq:rho1dot} is the crux of
1DMFT, which has
only very recently begun to be explored.  All the development
and results so far have been in the linear response
regime~\cite{PGB07,GBG08,PGGB07}, using functionals for $\rho_2$ which
have been adiabatically bootstrapped from ground-state
ones (eg.~Refs.\cite{BB02,SDLG08}).

The accuracy of the time-dependent PSDFT(1DMFT) depends on
approximate functionals for the term involving the second-order Wigner
function $w_2$($\rho_2$) in terms of $w$($\rho_1$). As has been noted
above, and earlier in the literature, the kinetic term is treated
exactly, in contrast to DFT, where the correlation potential/energy
has a kinetic component.
The simplest approximation would be the uncorrelated one, where $\rho_2$ is written as a product of antisymmetrized one-body $\rho_1$'s.
%\begin{equation}
%\rho_2({\bf{r_1}},{\bf{r'_1}};{\bf{r_2}},{\bf{r'_2}},t)=\rho_1({\bf{r_1}},{\bf{r'_1}},t) \rho_1({\bf{r_2}},{\bf{r'_2}},t)-\rho_1({\bf{r_1}},{\bf{r'_2}},t) \rho_1({\bf{r_2}},{\bf{r'_1}},t)
%\label{eq:uncorr}
%\end{equation}
This truncates the BBGKY heirarchy at the first level, and is equivalent to time-dependent Hartree-Fock. 
Its inability to change
occupation numbers makes it unable to treat many of the challenges that TDDFT faces, that the phase-space DFT would hope to treat. For example, 
the electronic quantum control
problem mentioned in the introduction (see also~\cite{AG09}). 
Section~\ref{sec:momentum} discusses another example where a change in occupation number is crucial for an accurate description. 
Natural orbitals $\psi_i(\br,t)$ with occupations $f_i(t)$ are defined by diagonalizing 
$\rho_1(\br,\br',t)$:
% into natural orbitals $\phi_i(\br,t)$ and occupation numbers $f_j(t)$:
\ben
\rho_1({\bf{r}},{\bf{r'}},t)=\sum_{i}f_i(t) \psi_i({\bf{r}},t) \psi^*_i({\bf{r'}},t)
\een
We may define ``natural Wigner orbitals'' via the analogous expansion:
\ben
w(\br,\bp,t) = \sum_{i}f_i(t)w_i(\br,\bp,t)
\een
where $w_i(\br,\bp,t) = \int d^3y\phi_i^*(\br -\by/2)\phi_i(\br+\by/2,t)$. 
From Eq.~\ref{eq:rho1dot} the time evolution of the occupation numbers can be derived:
\bea
i\dot{f_i}(t)=\int \int \int \frac{1}{|{\bf{r'-r_2}}|}\rho_2({\bf{r,r_2,r',r_2}},t)\nonumber\\
\phi^*_i({\bf{r}},t) \phi_i({\bf{r'}},t) d{\bf{r}}d{\bf{r'}}d{\bf{r_2}}-c.c
\eea
The right-hand-side of this equation is only nonzero if the contraction of
$\rho_2$ on to the natural orbitals, $\phi^*_i({\bf{r}},t)\rho_2({\bf{r,r_2,r',r_2}},t)\phi_i({\bf{r'}},t)$, is imaginary.
Therefore occupation numbers cannot change with using an uncorrelated product expression for $\rho_2$. It was also shown that bootstrapping any adiabatic functional from ground-state 1DMFT cannot change occupation numbers~\cite{PGGB07,heikothesis}. 
A systematic approach might consider then truncating the BBGKY heirarchy at
the next order; that is, solving the equation of motion for $\rho_2$ consistently 
with Eq.~\ref{eq:rho1dot}, but using the uncorrelated (antisymmetrized product form) for $\rho_3$. However, it has been shown that this violates fundamental 
trace relations between density matrices of different order~\cite{SRT90,CP92}; ways to get around this become rapidly complicated.
%There are ways to get around this...which become rapidly complicated;
Moreover, one must deal with a four-point function (second-order
Wigner function) rather than the two-point one (first-order).

The problem of an appropriate $w_2$-functional in terms of $w$, which
is able to change occupation numbers while retaining fundamental
physical properties, is an important area for future research.

%In terms of the natural orbitals, one can write $\rho_2$ as
%\begin{equation}
%\rho_2({\bf{r,r_2,r',r_2}},t)= 
%\\ 
%\sum_{i} c_{jklm}(t) \phi^*_j({\bf{r'}},t) \phi^*_k({\bf{r_2}},t) \phi_l({\bf{r}},t)\phi_m({\bf{r}},t)
%\end{equation}

%\subsection{One-to-one mapping between Wigner densities and the external potential}
%\label{sec:1-1}
%(**or put in an appendix?**) 
%The Runge-Gross (RG) theorem of TDDFT
%\cite{RG1} proves that for a specified initial state, there is a
%one-to-one mapping between the time-dependent coordinate-density and
%the external potential of the system. 
% Given that $w(\br,\bp,t)$
%yields a unique density $n(\br,t)$ (Eq.~\ref{eq:nrnp}), then
% a 1-1 mapping between $w$ and $v$ follows directly from
%invoking the RG theorem: 

\section{Revisiting two challenges within phase-space DFT}
\label{sec:examples}
We now describe in a little more detail two problems that are
particularly challenging for TDDFT approximations, and discuss how a PSDFT
can ameliorate them.

\subsection{Kohn-Sham momentum densities}
\label{sec:momentum}
In general, the momentum-distribution of the KS system is not the same
as that of the true system: although the sum of the squares of the KS
orbitals in coordinate-space yield the exact density of the true
system, the sum-square of their Fourier transforms to momentum-space
does not yield the exact momentum-density.  In cases where the
correlation component to the kinetic energy, $T\c$ is relatively small,  
calculating the
momentum distribution simply from the Fourier transforms of the
relevant KS orbitals can sometimes be quite accurate, as recently
illustrated by the Compton profiles obtained in Ref.~\cite{HZKP05},
provided accurate enough ground-state xc potentials are used.  Earlier, for the
the ground state case, Lam and Platzman derived a local-density
approximation for the correlation correction to KS
momentum-densities for Compton profiles~\cite{LP74}. 
The KS
orbital momentum distribution has also been successful in electron
momentum spectroscopy~\cite{HDCS02,DCCS94}: this is essentially a
triple differential cross-section measurement, where one electron is
scattered from an atom, molecule, or surface, causing the ejection of
another electron. The experiment obtains a map of the electron
momentum distribution before ejection, thus imaging the momentum
distribution of the Dyson orbital. Refs.~\cite{HDCS02,DCCS94} show
that the KS orbital yields a good approximation to the Dyson orbital
in their momentum distributions, better than, for example,
Hartree-Fock.

Away from the ground-state, Wilken and Bauer~\cite{WB07}
have shown that KS momentum-distributions of TDDFT are miserably poor for
non-sequential double-ionization processes; not only their shape is
wrong, lacking the characteristic double-hump structure, but also
their magnitude is significantly underestimated. Ref.~\cite{WB07} develops
a ``product-phase'' approximation to extract the true momentum distribution from the KS system, i.e. an observable functional for the momentum-distribution.

The possibility of working directly in momentum-space is hindered by
the lack of a one-to-one mapping theorem for momentum-densities (even
for the ground-state case)~\cite{DG99book,H81}.  In fact the
earlier interest in developing ways to associate phase-space
distributions to a given ground-state coordinate-space
density~\cite{PRG86,GP86,GBP84} was motivated by trying to get good
momentum-space properties derived from the coordinate-space density.
One may view the density-matrix developments as a rigorous approach to
this problem; as discussed in Sec.~\ref{sec:Wigner}, the exact one-body density matrix,
or equivalently, the one-body phase-space density, directly yields all
one-body information, i.e. the expectation value of any one-body
operator, such as the momentum-density operator, is exact.

%Because they can be experimentally verified by high-energy Compton
%scattering, single-particle momentum densities of many-particle
%systems are useful quantities to compute. Their importance lies in the
%fact they are relatively simple functions that incorporate the
%many-body aspects of the interactions between the particles of the
%system. However, the exact mapping between the time-dependent
%(single-particle) position density and the
%time-dependent-single-particle momentum density of a many body system
%is unknown. 

\begin{figure}[h]
\centering
\centerline{\psfig{file=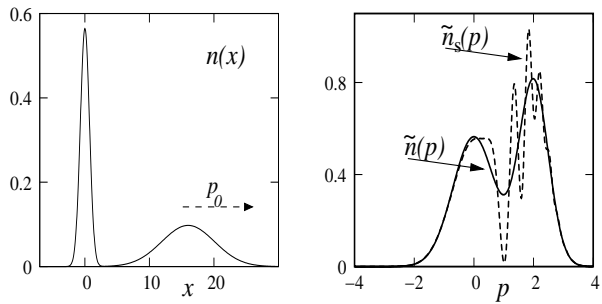,width=8cm,height=4cm,angle=0}}
\caption{Density profiles at time $t=8$, for parameters $\Delta = 0.8$ and $p_0 = 2$. The left-hand panel shows the coordinate-density profile, identical by definition for KS and true. The right-hand panel shows the true (solid) and KS (dashed) momentum-density profile.}
\label{fig:mom-dist}
\end{figure}

\begin{figure}[h]
%\centering
\centerline{\psfig{file=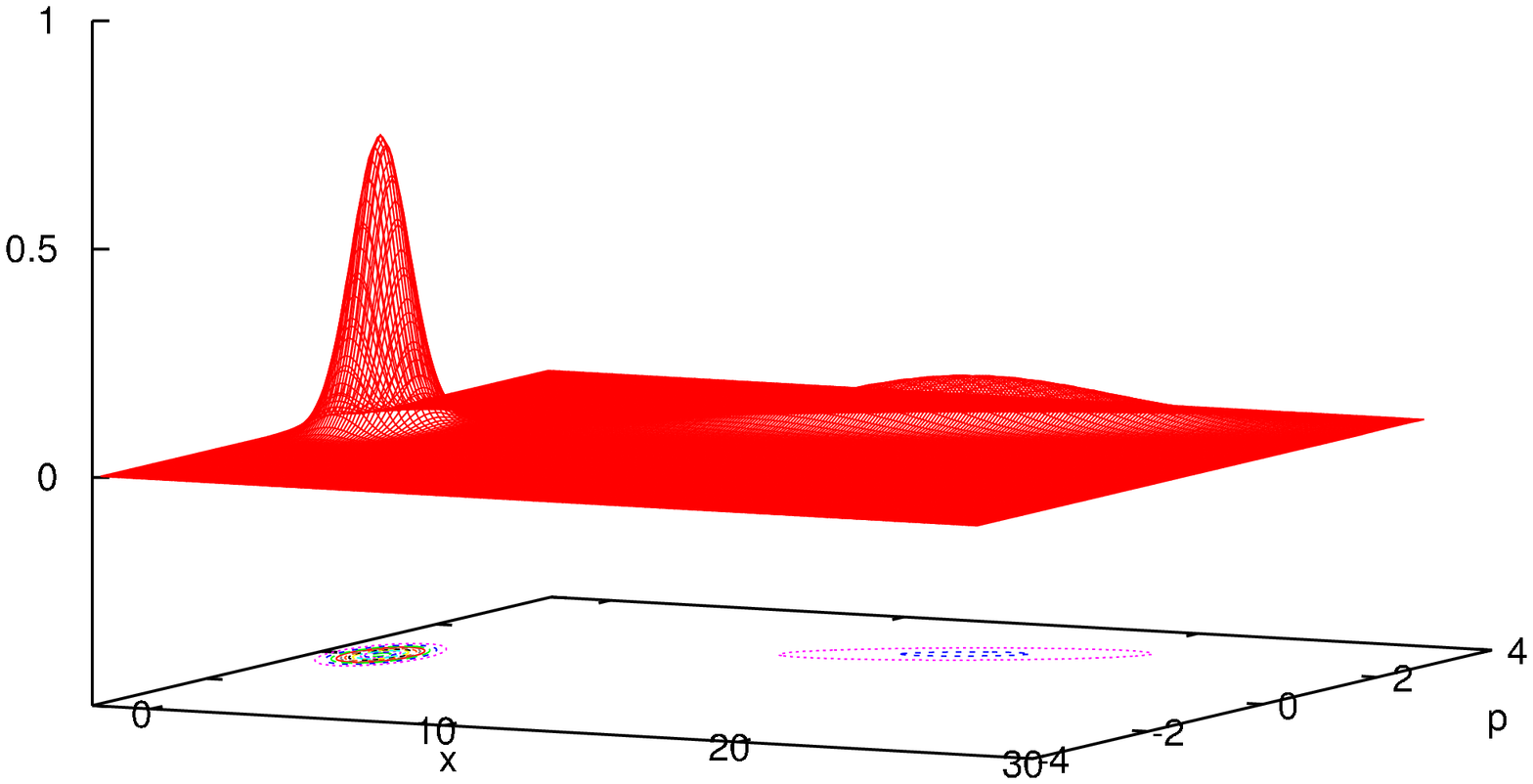,width=5cm,height=8cm,angle=270}}
\caption{The true Wigner profile at time $t=8$, for parameters $\Delta = 0.8$ and $p_0 = 2$.}
\label{fig:wigner_ion}
\end{figure}

Even a simple model consisting of two electrons in one-dimension
illustrates the discrepancy between the KS and true momentum-densities
in ionization or scattering problems. With an exactly solvable model,
one can pinpoint the origin of the differences between the momentum
distributions of the true and KS systems. In this particular example,
we will see it is due to the single-Slater determinant description
being a poor description of a state in which two electrons are doing
completely different things.  One electon is at rest (modelling a
``bound'' electron), while the other is moving away from it (modelling
a ``scattered'' or ``ionized''electron), and we consider that enough
time has elapsed that the true wavefunction describes independent
 electrons:
\ben
\Psi(x,x',t) = \left(\psi_0(x)\psi_p(x',t) + \psi_0(x')\psi_p(x,t)\right)/\sqrt{2}
\een
The orbital for the bound electron is simply chosen as a Gaussian
\begin{equation}
\psi_0(x)=e^{-x^2/2}/\pi^{\frac{1}{4}}
\end{equation} 
while that for the ionized electron is, for simplicity, chosen to be a free-particle dispersing Gaussian, which was initially ejected by some laser field (turned off thereafter), such that its initial spread in position is $1/\Delta$, and average momentum is $p_0$:
\begin{equation}
\psi_p(x,t)=\sqrt{\frac{\Delta}{\sqrt{\pi(1+it\Delta^2)}}}
e^{\left(\frac{ip_0x-\Delta^2x^2/2-ip_0^2t/2}{1+i\Delta^2t}\right)}
\end{equation}
The coordinate-density, $n(x,t)
= 2\int dx' \vert\Psi(x,x',t)\vert^2$, and momentum-density, $\tilde
n(p,t) = 2\int dp' \vert\int dxdx' e^{i(px+p'x')}\Psi(x,x',t)/(2\pi)\vert^2$ of this state at a snapshot in time is shown in
Fig.~\ref{fig:mom-dist}: Neither are remarkable, as both are what
might be expected quasiclassically from the sum of two probability
distributions centered around the average position and momentum for
each electron.  Now we consider the Kohn-Sham description, which, for
two electrons in a singlet state, operates via the doubly-occupied
orbital, reproducing the time-evolving density $n(x,t)$ of the true
system:
\begin{equation}
\phi(x,t)=\sqrt{\frac {n(x,t)}{2}} \exp\left(i\int^x dx' \frac{J(x',t)}{n(x',t)}\right) 
\end{equation}
where $J(x,t)$ is the current-density of the true system.  We compute
the KS momentum distribution, $\tilde n\s(p,t) =
\vert\tilde\phi(p,t)\vert^2$ from the Fourier transform $\tilde \phi$
of the orbital $\phi$. From Figure~\ref{fig:mom-dist}, we see that
while the overall envelope of the KS momentum distribution follows
that of the true system, there are distinct oscillations.  In
Figure~\ref{fig:wigner_ion} we plot the corresponding Wigner
phase-space distributions, $w(x,p,t)$ and $w_s(x,p,t)$; oscillations
are evident in the latter also, and dip into negative values.  This is
a typical signature of non-classical behavior: the KS orbital
describes an electron spatially delocalized in two regions, impossible
for a classical particle. The coherence between the two separated
parts of the wavepacket gives rise to the oscillations in the
phase-space density: integrating over momentum washes out these
oscillations, yielding the sum of the two Gaussians in the
left-hand-side of Fig.~\ref{fig:mom-dist} (i.e the true interacting
density) whereas the oscillations persist in the KS momentum
distribution on the right. On the other hand, the true Wigner function
looks much like a classical joint phase-space distribution would be: a
peak at the position and momentum of each electron, each locally
smeared out. Aside from the antisymmetrizing, the true wavefunction
has classical features, with each electron in their independent
Gaussian wavepackets, quite in contrast to the KS description which
describes both electrons in the same delocalized orbital. Although
yielding the exact coordinate density of the true system, this exact
KS wavefunction is clearly far from the true
wavefunction. Fundamentally, it associates to the true two-determinant
state, a {\it single} determinant; in a sense, this is a
time-dependent analog of the static correlation problems that haunt
ground-state density functional theory, for example in dissociation
problems~\cite{M05c,GB97,TMM09}.
We can understand the discrepancy in the momentum distributions as arising 
from this fundamental difference in the nature of the true and KS states in this
particular case. 
(In the next section, we shall see an example where the momentum distributions 
are not so different).
We stress that this is the KS system with the {\it exact} xc functional: what is required in TDDFT to get the correct exact momentum distribution from this, is
the (unknown) momentum observable functional.

\begin{figure}
%\centering
\centerline{\psfig{file=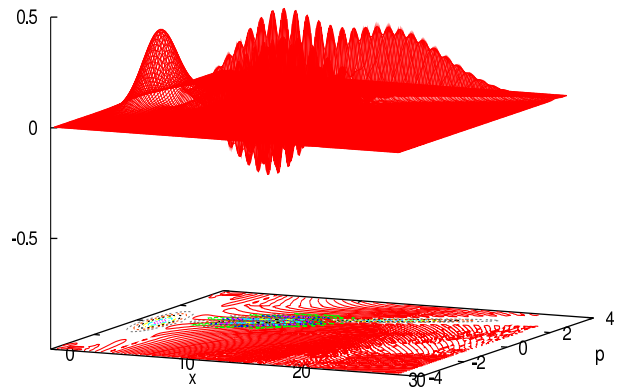,width=5cm,height=8cm,angle=270}}
\caption{The KS Wigner profile at time $t=8$, for parameters $\Delta = 0.8$ and $p_0 = 2$.}
\label{fig:wignerks_ion}
\end{figure} 

What are the implications of this for a PSDFT? As explained in the
earlier section, in a phase-space DFT we deal directly with the
interacting system; there is no non-interacting system that can
reproduce an interacting phase-space density. Momentum distributions
are immediately obtained without the need for an additional observable
functional. And, had we the exact functional for the $w_2$-term in
terms of $w$, these would be exact. In practise, approximations for
this term are needed, and for this example, it is crucial for the
approximation to be able to change occupation number. The state starts
in a ground-state, both electrons bound, which, in a first
approximation would be in the same doubly-occupied orbital.In the
process of ionization, one electron leaves, and a two-orbital description is necessary, each with occupation number 0.5. Any approximation unable to change occupation
numbers will once again yield wiggles in the momentum and phase-space
distribution. These should not be there in such a state which is of
classical nature.
We comment here that a similar problem occurs for certain quantum control
theory type problems, as discussed in the introduction.

\subsection{Dependence on the history of the density}
\label{sec:Hooke}
The time-dependent exchange-correlation potential functionally depends
on the initial state (both the true and KS) as well the history of the
density~\cite{RG84,MBW02,M05b}. This dependence on the past is
referred to as ``memory-dependence''. Any adiabatic approximation
(e.g., ALDA) utilizes only the instantaneous density, so ignore this
temporal nonlocality of the xc potential. Yet, adiabatic
approximations work remarkably well for the calculation of most, but
not all, excitations and their oscillator strengths. For dynamics in
strong fields, there are both parameter regimes and phenomena in which the memory-dependence is minimal, but also examples where memory-dependence is vital~\cite{M05b,MBW02,U06}. 
Memory-dependence is not easy to capture in approximations although there has been 
intense progress recently~\cite{DBG97,T05,T05b,UT06,KB05}.

Ref.~\cite{HMB02} used a numerically exactly-solvable example to
demonstrate explicitly a case where the correlation potential depends
on the density ultra-nonlocally in time. There, time slices were found
in which the density was practically identical, semi-locally in time,
whereas the correlation potential differed significantly. We shall
revisit this example here, with a view to asking whether functionals
in a PSDFT could be less non-local in time, and thus easier to
approximate. Specifically, we shall ask whether the {\it phase-space}
density can distinguish those time slices mentioned above, that the
coordinate-space density could not. We shall find that indeed it can,
and moreover, that the size of the differences in the phase-space
density tracks the difference in the correlation potential.  This
suggests that the memory-dependence in a PSDFT could be milder than in
the usual DFT.

\begin{figure}
%\begin{center}
\centering
\includegraphics[angle=0,width=9cm,height=6cm]{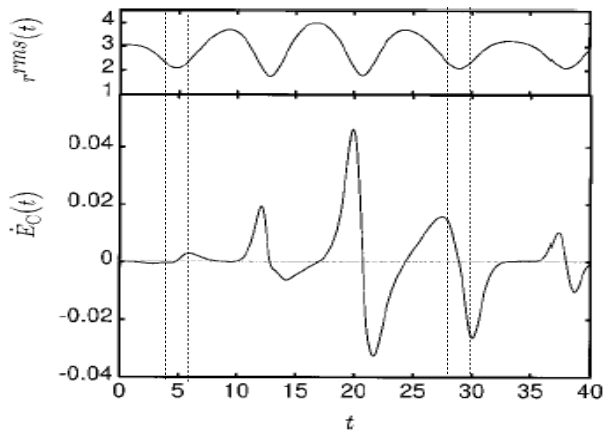}
\caption{Nonlocality of the correlation potential in time: $\int v\c(r,t)\dot n(r,t)d^3r$ (lower panel) and $r_{rms}(t)$ (top panel). In the two time slices indicated, the density is almost identical locally in time, whereas the density-weighted correlation potential in the lower panel is quite distinct.}
\label{fig:nonloc}
%\end{center}
\end{figure}

Time-dependent
Hooke's atom is a system of two Coulombically
interacting electrons in a time-dependent parabolic well: the 
Hamiltonian is
\begin{equation}
\hat{H}=-\frac{1}{2}\nabla^2_{1}-\nabla^2_{2}+\frac{1}{2}k(t)(r^2_1+r^2_2)+\frac{1}{|{\bf{r_1}}-{\bf{r_2}}|}
\end{equation}
For our investigations, we chose $k(t)=\overline{k} -\epsilon
\cos(\omega t)$, as in Ref.~\cite{HMB02}.  The Hamiltonian decouples
into relative ($u$) and center of mass ($R$) co-ordinates , so one may
solve for the exact time-dependent 2-electron wave function as a
product of $R$ and $u$ wavefunctions,
$\Psi(r_1,r_2,t)=\chi(R,t)\xi(u,t)$ and thereby obtain time-dependent
density $n({\bf{r}},t)$.  Knowledge of the time-evolution of the
density yields the doubly-occupied time-dependent KS orbital, and
then, via inversion of the time-dependent KS equation, the
exchange-correlation potential $v_{xc}(\br,t)$. A detailed
derivation can be found in Refs.~\cite{HMB02,HPB99}, along with computational details. We mention here only that the time-propagation method used is the Crank-Nicholson scheme, and a grid chosen to sample points near the origin more densely than further away, since the density decays exponentially. 
With chosen
parameters $\epsilon=0.1$, $\omega = 0.75$, and $\overline{k} = 0.25$,
the density retains a near-Gaussian shape at all times, with a
time-varying width; thus it may be parametrized by its variance
$r_{rms}(t) =\sqrt{\langle r^2\rangle}$~\cite{HMB02}.
This is shown in the top panel of the Fig.~\ref{fig:nonloc}. The lower
panel shows a density-weighted correlation potential, $\dot{E\c} =
\int d^3 r \dot n(\br,t) v\c(\br,t)$, which was also studied in the
earlier paper~\cite{HMB02}, as a simple way to track the
time-evolution of the correlation potential.  A pair of time
slices is indicated within which $r_{rms}$ is practically
identical in each time slice, while the density-weighted correlation
potential $\dot{E\c}$ differs quite dramatically. Other such pairs of time slices may be found. 
 This indicates that local, or even semi-local in time density-dependence is not enough to specify the correlation potential, i.e. that the correlation potential depends on a significant history of the density~\cite{HMB02,M05b}.
Any adiabatic approximation that uses only instantaneous
density information would erroneously predict the same correlation
potential for each time slice. 

We now consider the phase-space distribution at these time slices: if
these are also nearly identical at these pairs of time slices, then it
would suggest the memory-dependence in a PSDFT is just as non-local as
in DFT and just as challenging to model. We shall instead find that the phase-space distributions
distinguish the system at these different pairs of time slices, with
the differences greater with greater differences in the density-weighted
correlation potential $\dot E\c$. This suggests that perhaps the memory-dependence
would be less severe in PSDFT, and therefore perhaps easier to approximate.

\begin{figure}
%\begin{center}
\centering
\includegraphics[angle=0,width=8cm,height=6cm]{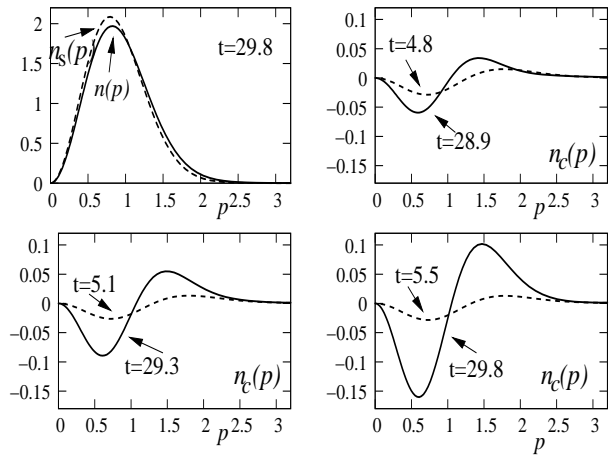}
\caption{Top left panel: the true (solid) and KS (dashed) momentum densities, at $t=29.8$. Top right and lower two panels:  
correlation components of the momentum distributions $n\c(p) $ at three different time-pairs indicated. In each time-pair, the coordinate density, $n(r,t)$, is identical, while the momentum-density is not, with the difference in its correlation component $n\c(p)$ growing as the difference in the density-weighted correlation potential $\dot{Ec}$ at those times.}
\label{fig:nc_hooke}
%\end{center}
\end{figure}

Knowledge of the true and KS wavefunctions, enables us to derive
expressions for Wigner phase-space distribution for the interacting
and KS cases:
\begin{eqnarray}
w({\bf{r}},{\bf{p}},t)= \left \vert \int_{0}^{\infty} x^2 dx \int_{-1}^{+1} e^{-ipx \cos\alpha \cos\theta}\right.\nonumber\\
\times \left.J_0(px \sin\alpha \sqrt{1-\cos^2\theta})\chi^*(\frac{1}{2} \sqrt{r_{+}},t) \xi(\sqrt{r_{-}},t) d(\cos\theta)\right \vert^{2}
\end{eqnarray}

\begin{eqnarray}
w\s({\bf{r}},{\bf{p}},t)=\int_{0}^{\infty} x^2 dx \int_{-1}^{+1} e^{-ipx \cos\alpha \cos\theta}\nonumber\\
\times J_0(2px \sin\alpha \sqrt{1-\cos^2\theta}) \phi^*(\frac{1}{2} \sqrt{r_{+}},t) \phi(\sqrt{r_{-}},t) d(\cos\theta)
\label{eq:ws}
\end{eqnarray}
where  $r_{+}$ = $r^2 + x^2 + 2rx \cos\theta$,  $r_{-}=r^2 +x^2 -2rx \cos\theta$ and  $\alpha$ is the angle between ${\bf{r}}$ and ${\bf{p}}$ and $J_0$ is the Bessel function of zeroth order. In Eq.~\ref{eq:ws}, $\phi$ is the KS orbital. 
Both distributions $w$ and $w\s$ depend only on the magnitudes of the coordinate and momentum, $r$ and $p$, and the phase-space angle $\alpha$; for simplicity, we look at the angle-averaged quantities:
\begin{equation}
w_{av}(r,p,t)=\int_{0}^{\pi} w(r,p,\alpha,t) \sin\alpha d\alpha
\end{equation}\\
Further, to highlight the component of the Wigner function that is due to correlation, we take the difference between the angle-averaged true and KS Wigner functions, defining
\ben
w\c(r,p,t)= w_{av}(r,p,t)- w_{s,av}(r,p,t)
\een
To simplify the analysis even further, we consider only the momentum-distribution of this correlation component, defined as:
\begin{equation}
\tilde n\c(p,t)= \tilde n(p,t) -\tilde n\s(p,t) = \int w_c(r,p,t) r^2 dr 
\end{equation}
We now turn to the plots in Figure~\ref{fig:nc_hooke}. 
The top left panel shows the true and KS momentum probability distributions at $t=29.8$, which has one of the larger values for the density-weighted correlation potential
in this run (Fig.~\ref{fig:nonloc}). We include in all these plots the Jacobian factor, so that we are actually plotting $p^2n(p)$. 
Although the momentum distributions are not
identical, the differences are relatively small, in contrast to
the ionization example in the previous section. 
%To study better these differences, we now turn to the other three panels.  
We now study their difference $n\c(p)$. 
The other three panels show $n\c(p)$ at pairs of times at which the coordinate-density are practically identical (i.e. essentially indistinguishable to the eye on the scale of plots similar to the top left panel). 
The first time-pair, (4.8,29.8), is chosen near the minimum of the $r_{rms}$ at each of the two time slices indicated in
 Figure~\ref{fig:nonloc}; here the difference in $\dot E\c$ is relatively small (0.0005 au), while it is bigger (0.011 au) at the second time pair (5.1,29.3), 
and bigger still (0.0264) at the third pair (5.5,29.3). 
In contrast to the
coordinate-density, the momentum-densities are not identical at these
pairs of times. Moreover, their difference increases with the
difference in the value of the density-weighted correlation potential at those times. 
The momentum distribution appears to somewhat track the correlation
potential. 

What are the implications of this for memory in a PSDFT? The results
suggest that having momentum information in the basic variable may
reduce their memory-dependence. Of course this is not conclusive, but
the fact that momentum distributions distinguish the system at pairs
of times when the coordinate-density does not, yet the correlation
potential is different, does suggest that functionals of the
phase-space density may be less non-local in time.

\section{Summary and Outlook}
A PSDFT approach could be more successful than the usual TDDFT for
some applications for which the usual functional approximations in
TDDFT perform poorly.  In many strong-field applications,
memory-dependence is required in the TDDFT xc potential, challenging
to model, and additional observable functionals are needed, depending
on the measured quantity of interest.  The examples worked through in
section~\ref{sec:examples} demonstrate each of these aspects
clearly. Being exactly-numerically solvable two-electron problems, one
can perform a detailed analysis that helps to understand these
challenges in their broader context.

 The first example, a model of ionization, illustrated, via a phase-space exploration, that even
 though an exact KS treatment gets the exact coordinate-space density
 correct, it may do so with a wavefunction that is fundamentally
 different than the true wavefunction; therefore approximating observables not directly related to the coordinate density by their value on the KS orbitals does not work well. We considered the momentum
 distribution (eg modelling that as measured by an ion-momentum recoil
 experiment), obtained from the KS orbitals and found spurious oscillations 
 as a consequence of the single-determinantal nature of the
KS wavefunction: a single delocalized orbital is describing the true
two-orbital state. The exact coordinate-space density is recovered (by
definition), but the momentum distribution displays oscillations,
present also in the KS phase-space distribution. The latter goes
negative, a sign of non-classical behavior: the KS system describes
the exact density putting both electrons in a doubly-occupied orbital,
which must therefore be delocalized in two regions. The true system is
however a rather classical one, aside from the antisymmetry, and this
is reflected in its phase-space density profile. 

The second example explored the memory-dependence of the correlation
potential in a time-dependent Hooke's atom. Solving directly for (a
density-weighted measure of) the correlation potential, time slices
were found in which the coordinate-density evolved identically, while
the correlation potential varied dramatically, illustrating the fact
that the correlation potential is a highly non-local in time
functional of the coordinate-density. We then asked whether, if
momentum-distribution information was included in the basic variable,
the correlation potential would be just as non-local in time as a
functional of the phase-space density, or whether the phase-space
density could distinguish the state of the system at these times. We
found the latter was true, suggesting that correlation functionals of
the phase-space density are less non-local in time, and therefore
easier to approximate than correlation functionals of the density
alone.
 
In PSDFT, the basic
variable is the one-body Wigner phase-space density, from which
momentum-space and position-space densities can be obtained via
integration over the conjugate variable.  Formally equivalent to
density-matrix functional theory, all one-body observables are
directly obtained from the basic variable (including the momentum distribution), without the need for additional observable
functionals. In its equation of motion the action of the kinetic
energy operator is exact as a phase-space functional (equally,
density-matrix functional), while approximations are needed for the
second-order Wigner function as a functional of the first-order
one. An important future direction is to develop functional
approximations for this that lead to changing occupation numbers, in
order to treat many of the problems that are challenging in the usual
TDDFT, such as electronic quantum control problems, and ionization
dynamics. In light of the studies in this paper, this would be well worth pursuing.

We gratefully acknowledge financial support from the National Science Foundation NSF CHE-0547913, and a Research Corporation Cottrell Scholar Award (NTM).

%We studied in some detail an exactly solvable two-electron problem
%that models the challenges that functionals in the usual TDDFT have in
%obtaining electron momentum distributions after ionization (eg. as
%measured in their ion's recoil momentum). Spurious oscillations emerge
%in the momentum distribution obtained directly from the exact KS
%orbitals,

\end{document}